\documentstyle[aps,preprint]{revtex}
\begin{document}
\draft
%%%%%%%%%%%%%%%%%%%%%%%%%%%%%%%%%%%%%%%%%%%%%%%%%%%%%%%%%%%%%%%%%%
%%%%%%%                                                     %%%%%%
%%%%%%%            DEFINITIONS                              %%%%%%
%%%%%%%                                                     %%%%%%
%%%%%%%%%%%%%%%%%%%%%%%%%%%%%%%%%%%%%%%%%%%%%%%%%%%%%%%%%%%%%%%%%%
%
\def\kln{\kappa_{L}^{NC}}
\def\krn{\kappa_{R}^{NC}}
\def\klc{\kappa_{L}^{CC}}
\def\krc{\kappa_{R}^{CC}}
\def\ttz{{\mbox {$t$-$t$-$Z$}}}
\def\bbz{{\mbox {$b$-$b$-$Z$}}}
\def\tta{{\mbox {$t$-$t$-$A$}}}
\def\bba{{\mbox {$b$-$b$-$A$}}}
\def\tbw{{\mbox {$t$-$b$-$W$}}}
\def\tltlz{{\mbox {$t_L$-$t_L$-$Z$}}}
\def\blblz{{\mbox {$b_L$-$b_L$-$Z$}}}
\def\brbrz{{\mbox {$b_R$-$b_R$-$Z$}}}
\def\tlblw{{\mbox {$t_L$-$b_L$-$W$}}}
\def\ppbar{\bar{{\rm p}}{\rm p}}
\def\beq{\begin{equation}}
\def\enq{\end{equation}}
\def\ra{\rightarrow}
%%%%%%%%%%%%%%%%%%%%%%%%%%%%%%%%%%%%%%%%%%%%%%%%%%%%%%%%%%%%%%%%%%%
%%%%%%%%%%%%%%%%%%%%%%%%%%%%%%%%%%%%%%%%%%%%%%%%%%%%%%%%%%%%%%%%%%%
%
\preprint{MSUHEP-94/06}
\title{
A GLOBAL ANALYSIS  OF THE TOP QUARK \\
COUPLINGS TO GAUGE BOSONS}
\author{ Ehab Malkawi and C.--P Yuan }
\address{
Department of Physics and Astronomy, Michigan State University \\
 East Lansing, MI 48824}
\date{\today}
\maketitle
\begin{abstract}
We propose to probe the electroweak symmetry breaking sector by
measuring the effective couplings of the top quark to gauge bosons.
Using precision LEP data, we constrain the non--universal couplings
of \ttz and \tbw, parameterized by $\kln$, $\krn$, $\klc$
and $\krc$, in the electroweak chiral lagrangian framework.
Different scenarios of electroweak symmetry breaking will
imply different correlations among these parameters.
We found that at the order of ${m_t}^{2}\log {\Lambda}^2$,
in which $\Lambda \sim 4 \pi v$ is the
cut--off scale of the effective theory,
$\kln$ is already constrained by LEP data.
In models with an approximate custodial symmetry, a positive $\klc$
is preferred. $\krc$ can be constrained by studying the direct
detection of the top quark at the Tevatron and the LHC. At the NLC,
$\kln$ and $\krn$ can be better measured.
\end{abstract}
\newpage
%\narrowtext
\widetext
\section{Introduction and Motivations}

    Despite the success of the Standard Model (SM) \cite{mele,alta},
there is little faith that the SM is the final theory.
The reasons behind this
are fundamental and basic \cite{pc}, {\it e.g.}, the SM contains many
arbitrary parameters with no apparent connections.
Besides,  the SM provides no
satisfactory explanation for the symmetry breaking mechanism which takes
place and gives rise to the observed mass spectrum of the gauge bosons
and fermions.
In this work, we study how to use the top quark to probe the origin of
the spontaneous symmetry breaking and the generation of fermion masses.

There are strong experimental and theoretical arguments
suggesting the top quark must exist \cite{kane}; {\it e.g.}, from the
measurement of the weak isospin quantum number of the left--handed $b$ quark
we know the top quark has to exist.
{}From the direct search at the Tevatron, assuming SM top quark,
$m_t$ has to be larger than 131\,GeV \cite{D0}.
Recently, data were presented by the CDF group at
FNAL to support the existence of a heavy top quark with mass
$m_t \sim 174 \pm 20 $\,GeV \cite{CDF}.
Furthermore, studies on radiative corrections concluded that
the mass ($m_t$) of a standard top quark
has to be less than $200$\,GeV \cite{mele}. However, there
are no compelling reasons to believe that the top quark couplings to light
particles should be of the SM nature.
Because the top quark is heavy
relative to other observed fundamental particles, one expects that any
underlying theory at high energy scale $\Lambda \gg m_t$ will
easily reveal itself at low energy through the effective interactions of
the top quark to other light particles.
Also because the top quark mass is of the order of the Fermi scale
$v={(\sqrt{2}G_F)}^{-1/2}=246$\,GeV, which characterizes the
electroweak symmetry breaking scale, the top quark may be a useful tool
to probe the symmetry breaking sector. Since the
fermion mass generation can be closely related to the
electroweak symmetry breaking,
one expects some residual
effects of this breaking to appear in accordance with the generated mass
\cite{pczh,sekh}. This means new effects should be more apparent in the
top quark sector than any other light sector of the theory.
Therefore, it is important to study the top quark system as a direct tool
to probe new physics effects \cite{kane}.

Undoubtedly, any real analysis including the top quark cannot be completed
without actually discovering it. In the SM, which is a renormalizable theory,
the couplings of the top quark to gauge bosons are fixed by the
linear realization of the gauge symmetry $SU(2)_{L}\times U(1)_Y$. However,
the top quark mass remains a free parameter in the theory (SM). If the top
quark is not a SM quark, then in addition to the unknown top mass,
the couplings of the top quark to gauge bosons are not known.
Also, the effective theory describing the top quark interactions at low
energy can be non--renormalizable. Therefore, to conclude
the properties of the top quark from
the radiative corrections is less vital
and predictive. Still, precision data at this stage are
our best hope to look for any possible deviation in the top quark sector from
the SM.

The goal of this paper is to study the couplings of the top quark to
gauge bosons from the precision data at LEP and examine how to improve
our knowledge about the top quark at the current and future colliders.
Also we will discuss how to use this knowledge to probe the symmetry
breaking mechanism.

   Generally one studies a specific model
({\it e.g.}, a grand unified theory)
valid up to some high energy scale and evolves that theory down
to the electroweak scale to compare its predictions with the precision
LEP data \cite{sekh,lopez,feld}.
Such an approach provides
a consistent analysis for low energy data. In addition to such a
model by model
study, one can incorporate new physics effects in a model independent
way formulated in terms of either
a set of variables \cite{pesk,bar,bar1,bar2}
or an effective lagrangian \cite{geor,gold,buch}. In this
work, we will adopt the latter approach. We simply address the
problem in the following way.
Assume there is an underlying theory at some high energy scale.
How does this theory {\it appreciably} manifest itself at low energy?
Because we do not know the shape of the underlying theory
and because a general treatment is usually very complicated,
we cannot provide a satisfying answer. Still, one can get some crude
answers to this question based on a few
{\it negotiable} arguments suggested by the status of low energy
data with the application of the electroweak chiral lagrangian.

 It is generally believed that new physics is likely to come in via processes
involving longitudinal gauge bosons (equivalent to Goldstone bosons)
and/or heavy fermions such as the top quark.
One commonly discussed method to probe the electroweak symmetry sector is
to study the interactions among the longitudinal
gauge bosons in the TeV
region. Tremendous work has been done in the literature \cite{wwww}.
However, this is not the subject of this paper.
As we argued above, the top quark plays an important role
in the search for new physics. Because of its heavy mass, new physics
will feel its presence easily and eventually may
show up in its couplings to the gauge bosons. If the top quark is
a participant in a dynamical symmetry breaking mechanism, {\it e.g.},
through the $\overline{t}t$ condensate (Top Mode Standard Model) \cite{tana}
which is suggested by the fact that its mass is of the order of the
Fermi scale $v$, then the top quark
is one of the best candidates for search of new physics.

An attempt to study the non--universal interactions of the top quark has
been carried out in Ref.~\cite{pczh} by Peccei {\it et al}. However,
in that study only the vertex \ttz was considered based on the
assumption that this is the only vertex which gains a significant
modification due to a speculated dependence of
the coupling strength on the fermion mass:
$\kappa_{ij} \leq {\cal O} \left ( \frac{\sqrt{m_{i}m_{j}}}{v} \right )$,
where $\kappa_{ij}$ parameterizes some new
dimensional--four interactions among gauge bosons and fermions $i$
and $j$. However, this is not the only possible pattern of interactions,
{\it e.g.}, in some Extended Technicolor models
\cite{sekh} one finds that the non--universal residual interactions
associated with the vertices \blblz, \tltlz and \tlblw
to be of the same order.
In section IV we discuss the case of the SM with a heavy Higgs boson
($m_H > m_t$) in which we find the size of the non--universal effective
interactions \tltlz and \tlblw
to be of the same order but with a negligible \blblz effect.

Here is the outline of our approach.
First, we use the chiral lagrangian approach \cite{wein,geor2,cole,fer}
to construct the most general $SU(2)_L \times U(1)_Y$ invariant
effective lagrangian including up to dimension--four operators
for the top and bottom quarks. Then
we deduce the SM (with and without a scalar Higgs boson) from this
lagrangian, and only consider new
physics effects which modify the top quark couplings to gauge
bosons and possibly the vertex \blblz.
With this in hand, we perform a comprehensive analysis using precision
data from LEP.
We include the contributions from the vertex \tbw in addition to the
vertex \ttz, and discuss the special
case of having a comparable size in \bbz as in \ttz.
Secondly, we build an  effective model
with an approximate custodial symmetry ($\rho \approx 1$)
connecting the \ttz and \tbw couplings.
This reduces the number of parameters in the effective
lagrangian and strengthens its structure and predictability.
After examining what we have learned from the LEP data, we study how
to improve our knowledge on these couplings at the SLC, the Tevatron, the
LHC (Large Hadron Collider) and the NLC (Next Linear Collider) \cite{nlc}.
(We use NLC to represent a generic $e^-e^+$ supercollider.)

The rest of this work is organized as follows. In section II
we provide a brief introduction to the chiral lagrangian
with an emphasis on the top quark sector. In section III
we present the complete analysis of the top quark interactions with gauge
bosons using LEP data for various scenarios of symmetry breaking
mechanism.
In section IV we discuss the heavy Higgs limit ($m_{H} > m_t$)
in the SM model as an example of our proposed effective model at the top
quark mass scale.
In section V we discuss how the SLC, Tevatron, LHC and NLC
can contribute to the measurement of these couplings.
Some discussion and conclusions are given in section VI.

\section{Introduction to the chiral lagrangian}

 The chiral lagrangian approach has been used
in understanding the low energy strong interactions because it can
systematically describe the phenomenon of spontaneous symmetry breaking
\cite{wein}. Recently, the chiral lagrangian technique has been widely
used in studying the electroweak sector
\cite{gold,fer,pecc,hol,how1,fer1,app},
to which this work has been directed.

The chiral lagrangian can be constructed solely
on symmetry with no
other assumptions regarding explicit dynamics. Thus,
it is the most general effective lagrangian
that can accommodate any truly fundamental theory possessing that
symmetry at low energy. Since one is interested in the low energy
behavior of such a theory, an expansion in powers of the external momentum is
performed in the chiral lagrangian \cite{geor2}.

  In general one starts from a Lie group G which breaks down spontaneously
into a subgroup H, hence a Goldstone boson for every broken
 generator is to be introduced \cite{cole}.
 Consider for example G~$=SU(2)_L\times U(1)_Y$ and H~$=U(1)_{em}$.
There are three Goldstone bosons generated by this breakdown,
$\phi^{a},\,a=1,2,3$ which are eventually eaten by $W^{\pm}$ and $Z$
and become the longitudinal degree of freedom of these gauge bosons.

 The Goldstone bosons transform non--linearly under G but linearly
under the subgroup H. A convenient way to handle this is to introduce
the matrix field
\beq
\Sigma =exp(i\frac{\phi^{a}\tau^{a}}{v_{a}})\, ,
\enq
where $\tau^{a},\, a=1,2,3 $ are the Pauli matrices normalized as
${\rm{Tr}}(\tau^a \tau^b)=2 \delta_{ab}$. Because of $U(1)_{em}$ invariance
$v_1=v_2=v$, but is not necessarily equal to $v_3$.
The matrix field $\Sigma$ transforms under G as
\beq
\Sigma\rightarrow {\Sigma}^{\prime}=exp(i\frac{\alpha^{a}\tau^{a}}{2})\,
\Sigma \,exp(-iy\frac{\tau^3}{2})\, ,
\enq
where $\alpha^{1,2,3}$ and $y$ are
the group parameters of G.

In the SM, being a special case of the chiral lagrangian,
$v=246$\,GeV is the vacuum expectation value of the Higgs
boson field. Also
$v_3=v$ arises from the approximate custodial symmetry
in the SM.
It is this symmetry that is responsible for the
tree level relation
\beq
\rho= \frac{M_W}{M_Z\,\cos \theta_W}=1\,
\label{yeq3}
\enq
in the SM, where $\theta_W$ is the electroweak mixing angle.
In this work, we assume the full theory guarantees that
$v_1=v_2=v_3=v$.

Out of the Goldstone bosons and the gauge boson fields one can construct
the bosonic gauge invariant terms in the chiral lagrangian
\beq
{\cal L}_{B} =-\frac{1}{4} {W_{\mu \nu}}^{a} {W^{\mu \nu}}^{a}
 -\frac{1}{4} B_{\mu \nu}B^{\mu \nu}
 +\frac{1}{4}v^{2}\rm{Tr}({D_{\mu}\Sigma}^{\dagger}D^{\mu}
\Sigma)\, ,
\label{yeq4}
\enq
where the covariant derivative
\beq
D_{\mu}\Sigma =\partial_{\mu}\Sigma - ig{W_{\mu}}^{a}\frac{\tau^{a}}{2}\Sigma
+i{g}^{\prime}\Sigma B_{\mu}\frac{\tau^{3}}{2}\, .
\enq
In the unitary gauge $\Sigma =1$, one can easily see how the gauge bosons
acquire a mass.
In Eq.~(\ref{yeq3}), $M_W=gv/2$ is the mass of $W^\pm_\mu=
(W^1_\mu\mp i W^2_\mu)/\sqrt{2})$, $M_Z=gv/2/\cos \theta_W$
is the mass of $Z_\mu=\cos \theta_W W^3_\mu - \sin \theta_W B_\mu$.
The photon field will be denoted as
$A_\mu = \sin \theta_W W^3_\mu + \cos \theta_W B_\mu$.

Fermions can be included in this context by assuming that they
transform under G$ =SU(2)_L\times U(1)_{Y}$ as \cite{pecc}
\beq f\rightarrow {f}^{\prime}=e^{iyQ_f}f \label{eq1} \, ,
\enq
where $Q_{f}$ is the electromagnetic charge of $f$.

 Out of the fermion fields $f_1$, $f_2$ and the Goldstone bosons matrix
field $\Sigma$
the usual linearly realized fields
$\Psi$ can be constructed. For example, the left--handed
fermions ($SU(2)_L$ doublet) are constructed as
\beq
\Psi_{L} = \Sigma F_{L} = \Sigma{f_1\choose {f_2}}_{L} \label{psi} \,
\enq
with $Q_{f_1}-Q_{{f_2}}=1$.
One can easily show that $\Psi_{L}$\,transforms under G linearly as
\beq
\Psi_{L}\rightarrow {{\Psi}^{\prime}}_{L}=g\Psi_{L}\, ,
\enq
where $ g=
exp(i\frac{\alpha^{a}\tau^{a}}{2})exp(i\frac{y}{2})
 \in {\rm{G}} $.
Linearly realized right--handed fermions
$\Psi_{R}$  ($SU(2)_L$ singlet) simply coincide with $F_{R}$, {\it i.e.},
\beq
\Psi_{R}= F_{R}={f_1\choose {f_2}}_{R}\, .\label{psr}
\enq
Out of those fields with the specified transformations it is straightforward
to construct
a lagrangian which is invariant under $SU(2)_L\times U(1)_Y$.

Since the interactions among the light fermions and the gauge bosons have
been well tested to agree with the SM, we only consider
new interactions involving
the top and bottom quarks. We ignore all possible mixing of the
top quark with light fermions in these new interactions.
In case there exists a fourth generation with heavy
fermions, there can be a substantial
impact on the Cabibbo--Kobayashi--Maskawa
matrix element $V_{tb}$.
To be discussed below, this effect is effectively included
in the new non--standard couplings of \tbw.

Following Ref.~\cite{pecc}, define
\beq
{\Sigma_{\mu}}^{a}=-\frac{i}{2}{\rm{Tr}}(\tau^{a}\Sigma^{\dagger}D_{\mu}
\Sigma) \, ,
\enq
which transforms under G as:
\beq
 {\Sigma_{\mu}}^{3}\rightarrow {{{\Sigma}^{\prime}}_{\mu}}^{3}
       ={\Sigma_{\mu}}^{3}\, ,
\enq
\beq
{\Sigma_{\mu}}^{\pm}\rightarrow {{{\Sigma}^{\prime}}_{\mu}}^{\pm}
   =e^{\pm iy}{\Sigma_{\mu}}^{\pm}\, ,
\enq
where,
\beq
{\Sigma_{\mu}}^{\pm}=\frac{1}{\sqrt{2}}({\Sigma_{\mu}}^{1}\mp i
{\Sigma_{\mu}}^{2})\, .
\enq
In the unitary gauge $\Sigma =1$, we have
\beq
{\Sigma_{\mu}}^{3}= -\frac{1}{2}\frac{gZ_{\mu}}{\cos{\theta_W}}\, ,
\enq
\beq
{\Sigma_{\mu}}^{\pm}= -\frac{1}{2}g{W_{\mu}}^{\pm}\, .
\enq

Consider the interaction terms up to dimension--four for the
$t$ and $b$ quarks.
{}From Eqs.~(\ref{psi}) and (\ref{psr}), we denote
\beq
F= {t\choose b} = F_L + F_R\, ,
\enq
with $f_1=t$ and $f_2=b$.
The SM lagrangian can be deduced from
\begin{eqnarray}
{\cal L}_{0}&=&\overline{F}i\gamma^{\mu} \left ( \partial_{\mu}
-ig^{\prime} (\frac{Y}{2}+\frac{\tau^3}{2})
B_{\mu} \right ) F - \overline{F}\,M\,F \nonumber \\
&-& \overline{F_{L}}\gamma^{\mu}\tau^{a} F_{L}{\Sigma_{\mu}}^{a}
+{\cal L}_{B} \, \, ,
\end{eqnarray}
where $Y=1/3$ and $M$ is a diagonal mass matrix
\beq
 M=\pmatrix{m_t & 0 \cr 0 & m_b \cr}\, .
\enq
${\cal L}_{0}$ is invariant under G, and the electric charge of fermions
is given by $Y/2+T^3$, where $T^3$ is the weak isospin quantum number.
Taking advantage of the chiral lagrangian approach,
additional non--standard interaction terms, invariant under G, are allowed
\cite{pecc}
\begin{eqnarray}
{\cal L}&=& -\kln \overline{t_{L}}\gamma^{\mu} t_{L}{\Sigma_{\mu}}^{3}
  -\krn \overline{{t}_{R}}\gamma^{\mu} t_{R}{\Sigma_{\mu}}^{3} \nonumber \\
&-&\sqrt{2}\klc \overline{{t}_{L}}\gamma^{\mu} b_{L}{\Sigma_{\mu}}^{+}
-\sqrt{2}{\klc}^{\dagger}\overline{{b}_{L}}\gamma^{\mu}t_{L}
{\Sigma_{\mu}}^{-} \nonumber \\
&-&\sqrt{2}\krc \overline{{t}_{R}}\gamma^{\mu} b_{R}{\Sigma_{\mu}}^{+}
-\sqrt{2}{\krc}^{\dagger} \overline{{b}_{R}}\gamma^{\mu} t_{R}
{\Sigma_{\mu}}^{-} \label{eq2} \, ,
\end{eqnarray}
where $\kln$, $\krn$ are two arbitrary real parameters,
$\klc$, $\krc$ are two arbitrary complex parameters and the superscript
$NC$ and $CC$ denote neutral and charged current, respectively.
In the unitary gauge we derive the following non--standard terms
in the chiral lagrangian with the symmetry
$ { {SU(2)_L \times U(1)_Y} \over {U(1)_{em}} }$
\begin{eqnarray}
{\cal L}&=&\frac{g}{4\cos \theta_W}\bar{t}\left ( \kln \gamma^{\mu}
(1-\gamma_{5})+\krn \gamma^{\mu}(1+\gamma_{5})
\right ) t\,Z_{\mu} \nonumber \\
 &+&\frac{g}{2\sqrt{2}}\bar{t}\left ( \klc\gamma^{\mu}(1-\gamma_{5})
+ \krc \gamma^{\mu}(1+\gamma_{5})\right ) b\,{W_{\mu}}^{+} \nonumber \\
&+&\frac{g}{2\sqrt{2}}\bar{b}\left ( {\klc}^{\dagger}\gamma^{\mu}
(1-\gamma_{5})+{\krc}^{\dagger}\gamma^{\mu}(1+\gamma_{5}) \right )
t\,{W_{\mu}}^{-} \label{eq3} \, .
\end{eqnarray}

A few remarks are in order regarding the lagrangian
${\cal L}$ in Eqs. (\ref{eq2}) and (\ref{eq3}).
\begin{itemize}
\item[1)]
In principle, ${\cal L}$ can include
non-standard neutral currents $\overline{b_{L}}\gamma_{\mu}b_{L}$ and
$\overline{b_{R}}\gamma_{\mu}b_{R}$.
For the left-handed neutral current $\overline{b_{L}}\gamma_{\mu}b_{L}$
we discuss two cases: \\
a) The effective left-handed vertices \tltlz,
\tlblw and \blblz are comparable
in size as in some Extended Technicolor models \cite{sekh}.
In this case, the top quark contribution to low energy observables
is of higher order through radiative corrections,
therefore its contribution will be suppressed by $1/16{\pi}^{2}$.
In this case, as we will discuss in the next section,
the constraints derived from low energy data on the
non--standard couplings are so stringent (of the order of a few percent)
 that it would be a challenge to directly probe the non--standard
top quark couplings at the Tevatron, the LHC and the NLC.\\
b) The effective left--handed vertex \blblz is small
as compared to the
\ttz and \tbw vertices.
We will devote most of this work to the case where the vertex \blblz
is not modified by the dynamics of
the symmetry breaking.
This assumption leads to interesting conclusions to be seen
in the next section. In this case one needs to consider
the contributions of the top quark to low energy data through loop
effects. A specific model with such properties is given in section IV.
\item[2)]
We shall assume that \brbrz is not modified by the
dynamics of the electroweak symmetry breaking. This is the case in
the Extended
Technicolor models discussed in Ref. \cite{sekh}.
The model discussed in section IV is another example.
\item[3)]
The right--handed charged current contribution
$\krc$ in Eqs.~(\ref{eq2}) and (\ref{eq3}) is expected to be
suppressed by the bottom quark mass.
This can be understood in the following way. If $b$ is
massless ($m_b=0$), then
the left-- and right--handed $b$ fields can be
associated with different global $U(1)$ quantum numbers. ($U(1)$ is
a chiral group, not the
hypercharge group.) Since the underlying theory has an exact
$SU(2)_L\times U(1)_Y$ symmetry at high energy,
the charged currents are purely left--handed before the
symmetry is broken.
After the symmetry is spontaneously
broken and for a massless $b$ the $U(1)$ symmetry associated
with $b_{R}$ remains exact (chiral invariant) so it is not possible to
generate right--handed charged currents.
Thus $\krc$ is usually suppressed by the bottom quark mass although
it could be enhanced in some
models with a larger group G, {\it i.e.}, in models
containing additional right--handed gauge bosons.

We find that in the limit of ignoring the bottom quark mass,
$\krc$ does not contribute to low energy data through loop insertion
at the order ${m_t}^{2}\log {\Lambda}^2$, therefore we can not constrain
$\krc$ from the LEP data.
 However, at the Tevatron and the LHC $\krc$ can be measured
by studying the direct detection of the top quark
and its decays.
This will be discussed in {\mbox {section V.}}
\end{itemize}

It is worth mentioning that the photon does not participate in
the new non--universal interactions as described in the chiral lagrangian
${\cal L}$ in Eq.~(\ref{eq3}) because the
$U(1)_{em}$ symmetry remains an exact symmetry of the effective
theory.
Using Ward identities one can show
that such non--universal terms should not
appear. To be precise, any new physics can only contribute to the
universal interactions of the photon to charged fields. This effect
can simply be absorbed in redefining  the
electromagnetic fine structure constant $\alpha$, hence no new
\tta or \bba interaction terms
will appear in the effective lagrangian after a proper renormalization of
$\alpha$.

Here is a final note regarding the physical Higgs boson.
It is known that the gauge bosons acquire masses through the spontaneous
symmetry breaking mechanism. In the chiral lagrangian
this can be seen from the last term in ${\cal L}_B$ (see Eq.~(\ref{yeq4})),
which only involves the gauge bosons and the unphysical Goldstone bosons.
This indicates that the chiral lagrangian can account for the mass generation
of the gauge bosons without the actual details of the symmetry breaking
mechanism. Furthermore,
 the fermion mass term is also allowed in the chiral lagrangian,
\beq
 -m_{f_i}\overline{f_i}f_i \, , \nonumber
\enq
because it is invariant under G, where
the fermion field $f_i$ transforms as
in Eq. (\ref{eq1}).

{}From this it is clear the Higgs boson is not necessary in constructing the
low energy effective lagrangian. Indicating that the SM Higgs mechanism
is just one example of the possible
spontaneous symmetry breaking scenarios which might take place in nature.
Still, a Higgs boson
can be inserted in the chiral lagrangian as an additional field
($SU(2)_{L}\times U(1)_Y$ singlet) with arbitrary
couplings to the rest of the fields. To retrieve the SM Higgs boson
contribution at tree level, one can
 simply substitute the fermion mass $m_f$
by $g_f v$ and $v$ by $v+H$, where $g_f$ is the Yukawa coupling for fermion
$f$ and $H$ is the Higgs boson field.
Hence, we get the scalar sector lagrangian
\beq
{\cal L}_{H}=\frac{1}{2}\partial_{\mu} H \partial^{\mu} H
-\frac{1}{2}m_H H^2 -V(H) +\frac{1}{2}vH{\rm{Tr}}
\left ( {D_{\mu}\Sigma}^{\dagger}D^{\mu} \Sigma \right )
+\frac{1}{4}H^2{\rm{Tr}}
\left ( {D_{\mu}\Sigma}^{\dagger}D^{\mu} \Sigma \right )\, ,\label{higg}
\enq
where $V(H)$ describes the Higgs boson self--interaction.
The coefficients of the last two terms in the above equation can be
arbitrary for a chiral lagrangian with a scalar field other than
the SM Higgs boson.

In this analysis, we will discuss models
with and without a Higgs boson.
In the case of a light Higgs boson ($m_H <  m_t$) we will include the
Higgs boson field in the chiral lagrangian as a
part of the light fields with no new physics being associated with it.
In the case of a heavy Higgs boson ($m_H > m_t$) in the full theory,
we assume the Higgs boson field has been integrated out and its effect
on low
energy physics can be thought of as a new heavy physics effect which is
already included in the effective couplings of the top quark
at the scale of $m_t$. Finally, we
will consider the possibility of a spontaneous symmetry breaking
scenario without including a SM Higgs boson in the full theory.
In this case we consider the effects on low energy data from the
new physics parameterized by the non--standard
interaction terms in ${\cal L}$ in Eq.~(\ref{eq3})
and contributions from the SM
without a Higgs boson.

\section{ the Top Quark Couplings to Gauge Bosons}

As we discussed in the previous section, one possibility of new physics
effects is the modification of the vertices \bbz, \ttz and \tbw in the
effective lagrangian
by the same order of magnitude
\cite{sekh}. In this case, only the vertex \bbz can have large
contributions to low
energy data while, based on the dimensional counting method,
the contributions from the other two vertices
\ttz and \tbw are suppressed by
$1/16{\pi}^2$ due to their insertion in loops.

In this case, one can use $\Gamma_b$ (the partial decay width of the
$Z$ boson to $\overline{b}b$) to constrain the \bbz coupling.
Denote the non--standard \bbz vertex as
\beq
\frac{g}{4\cos \theta_W}\kappa \gamma_{\mu}(1-\gamma_5)\, ,
\enq
which is purely left-handed.
In some Extended
technicolor models, discussed in Ref.~\cite{sekh},
this non--standard effect arises from the same source as the mass generation
of the top quark, therefore
 $\kappa$ depends on the top quark mass.

As we will discuss later, the non--universal contribution to $\Gamma_b$
is parameterized by a measurable parameter denoted as
$\epsilon_b$ \cite{bar,bar1,bar2} which is measured to be \cite{bar}
\beq
\epsilon_b\,(10^{3}) = 4.4 \pm 7.0 \,. \nonumber
\enq
The SM contribution to $\epsilon_b$ is calculated in Refs.~\cite{bar,bar1},
{\it e.g.}, for a 150\,GeV top quark
\beq
{\epsilon_b}^{SM}\, (10^{3}) =-4.88 \,\,. \nonumber
\enq
The contribution from $\kappa$ to $\epsilon_b$ is
\beq
\epsilon_b=-\kappa \, .
\enq
Within 95\% confidence level (C.L.), from  $\epsilon_b$ we find that
\beq
-22.9 \leq \kappa\, (10^{3}) \leq 4.4 \, .
\enq

As an example, the simple commuting Extended Technicolor model
presented in {\mbox Ref. \cite{sekh}} predicts that
\beq
\kappa \approx  \frac{1}{2}{\xi}^2 \frac{m_t}{4\pi v} \, ,
\enq
where $\xi$ is of order 1.
Also in that model the top quark couplings
$\kln$, $\krn$ and $\klc$, as defined in Eqs.~(\ref{eq2}) and (\ref{eq3}),
are of the same order as $\kappa$.
For a 150\,GeV top quark, this model predicts
\beq
\kappa \, (10^{3})\approx 24.3\, {\xi}^2 \, \nonumber .
\enq
Hence, such a model is likely to be excluded using low energy data.

We will devote the subsequent discussion to models in which the
non--standard \bbz coupling can be ignored relative to the
\ttz and \tbw couplings. In this
case one needs to study their effects at the quantum level, {\it i.e.},
through loop insertion. We will first discuss the general case where no
relations between the couplings are assumed. Later we will
impose a relation between $\kln$ and $\klc$ which are defined in
Eqs.~(\ref{eq2}) and (\ref{eq3}) using an effective model with an
approximate custodial symmetry.

\subsection{General case}

  The chiral lagrangian in general has a complicated structure and many
arbitrary coefficients which weaken its predictive power. Still, with
a few further assumptions, based on the status of
present low energy data, the chiral lagrangian can provide a useful
approach to confine the coefficients parameterizing new physics effects.

In this subsection, we provide a general treatment for
the case under study with
minimal imposed assumptions in the chiral lagrangian. In this case, we
only impose the assumption that the vertex \bbz is not modified by
the dynamics. In the chiral lagrangian
${\cal L}$, as defined in Eqs.~(\ref{eq2}) and (\ref{eq3}),
there are six independent parameters ({\mbox {$\kappa$'s}})
which need to be constrained
using precision data.
Throughout this paper, we will only consider the
insertion of {\mbox {$\kappa$'s}} once in one--loop diagrams by assuming that
these non--standard couplings are small;
$\kappa_{L,R}^{NC,CC} < {\cal O}(1)$.
At the one--loop level
the imaginary parts of the couplings do not contribute to those
LEP observables of interest.
Thus, hereafter we drop the imaginary pieces from the effective couplings,
which reduces the number of relevant parameters to four.
Since the bottom quark
mass is small relative to the top quark mass,
we find that  $\krc$ does not contribute
to low energy data up to the order ${m_t}^2\log {\Lambda}^2$
in the $m_{b}\ra 0$ limit.
With these observations we conclude that only the three parameters
$\kln$, $\krn$ and $\klc$ can be constrained.

A systematic approach can be implemented for such an
analysis based on the scheme used in Refs.~\cite{bar,bar1,bar2}, where the
radiative corrections can be parameterized by 4 independent parameters,
three of those parameters $\epsilon_1$, $\epsilon_2$, and $\epsilon_3$
are proportional to the variables $S$, $U$ and $T$ \cite{pesk},
and the fourth  one; $\epsilon_b$ is due to the GIM violating
contribution in $Z\rightarrow b \overline{b}$ \cite{bar}.

  These parameters are derived from four basic measured \mbox{observables},
$\Gamma_{\ell}$\,(the partial width of $Z$ to a
charged lepton pair),
$A_{FB}^{\ell}$\,(the forward--backward asymmetry at the $Z$ peak for
the charged lepton $\ell$), $ M_{W} / M_{Z} $
and $\Gamma_{b}$\,(the partial width
of $Z$ to a $b\overline{b}$ pair).
The expressions of these observables in terms of
$\epsilon$'s are given in Refs.~\cite{bar,bar1}.
In this paper we only give the relevant terms in $\epsilon$~'s
which might contain the leading effects from new physics.

We denote the vacuum polarization for the $W^1, W^2, W^3, B$ gauge bosons as
\beq
{{\Pi}^{ij}}_{\mu \nu}(q) = -ig_{\mu \nu}\left [A^{ij}(0) + q^{2}F^{ij}(
q^2)
\right ] + q_{\mu}q_{\nu}\,\rm{terms}\, ,
\enq
where $i,j=1,2,3,0$ for $W^1, W^2, W^3$ and $B$,  respectively. Therefore,
\beq
\epsilon_1 = e_1-e_5 \,,
\enq
\beq
\epsilon_2 = e_2 - c^{2}e_5\, ,
\enq
\beq
\epsilon_3 = e_3 - c^{2}e_5 \, ,
\enq
\beq
\epsilon_b = e_b \, ,
\enq
where
\beq
e_1 = \frac{A^{33}(0) - A^{11}(0)}{{M_W}^{2}}\, ,
\enq
\beq
e_2 = F^{11}({M_W}^2) - F^{33}({M_Z}^2)\, ,
\enq
\beq
e_3 = \frac{c}{s}F^{30}({M_Z}^2)\, ,
\enq
\beq
e_5 = {M_Z}^{2}\frac{dF^{ZZ}}{dq^2}({M_Z}^2)\, ,
\enq
and $c\equiv \cos \theta_W$.
\beq
c^{2} \equiv \frac{1}{2}\left [ 1+{\left ( 1-\frac{4\pi \alpha (M_Z)}
{\sqrt{2} G_f {M_Z}^2}\right )}^{1/2} \right ] \, ,
\enq
and $s^2=1-c^2$.
$e_b$ is defined through the GIM-violating $Z\rightarrow b\overline{b}$
vertex
\beq
{V_{\mu}}^{GIM}\left ( Z \rightarrow b\bar{b}\right ) = -\frac{g}{2c}e_b
\gamma_{\mu}\frac{1-\gamma_5}{2}\, .
\enq

$\epsilon_1$ depends quadratically on $m_t$ \cite{bar,bar1}
and has been measured to
good accuracy, therefore $\epsilon_1$ is sensitive to any
new physics coming through the top quark. On the contrary, $\epsilon_2$\, and
$\epsilon_3$ do not play any
significant role in our analysis because their dependence on the top mass is
only logarithmic.

 Non--renormalizability of the effective lagrangian presents
a major issue of how to find a scheme to
handle both the divergent and the finite pieces in
loop calculations \cite{burg,mart}. Such a problem arises because one
does not know the underlying theory, hence no matching can be performed
to extract the correct scheme to be used in the effective lagrangian
\cite{geor}.
One approach is to associate the divergent piece in
loop calculations with a physical
cut--off $\Lambda$, the upper
scale at which the effective lagrangian is
valid \cite{pecc}. In the chiral lagrangian approach this cut--off
$\Lambda$ is taken to be
$4\pi v \sim 3$\,TeV \cite{geor}. For the finite piece no
completely satisfactory approach is available \cite{burg}.

To perform calculations using the chiral lagrangian,
one should arrange
the contributions in powers of $1/4\pi v$ and then include all
diagrams up to the desired power. In
the $R_{\xi}$ gauge ($\Sigma \neq 1$), the couplings of
the Goldstone bosons to the fermions should also be included
in Feynman diagram calculations.
These couplings can be easily found
by expanding the terms in ${\cal L}$ as given
in Eq.~(\ref{eq2}).
We will not give the explicit expressions for those terms here.
Some of the relevant Feynman diagrams are shown in Fig.~(\ref{fey}).
Calculations were done in the
't Hooft--Feynman gauge.
We have also checked our calculations in both the Landau gauge and the
unitary gauge and found agreement as expected.

We calculate the  contribution to $\epsilon_1$ and
$\epsilon_b$ due to the new interaction terms in the chiral lagrangian
(see Eqs.~(\ref{eq2}) and (\ref{eq3})) using the dimensional regularization
scheme and
taking the bottom mass to be zero.
At the end of the calculation, we
replace the divergent piece $1/\epsilon$ by
$\log(\Lambda^2/{m_t}^2)$ for $\epsilon = (4-n)/2$ where $n$ is the
space--time dimension. Since we are mainly interested
in new physics associated with the top quark couplings to gauge bosons,
we shall restrict ourselves
to the {\it leading} contribution enhanced by the top quark mass, {\it i.e.},
of the oder of ${m_t}^{2}\log {\Lambda}^{2}$.

We find
\beq
\epsilon_1=\frac{G_F}{2\sqrt{2}{\pi}^2}3{m_t}^2
 (-\kln+\krn+\klc)\log{\frac{{\Lambda}^2}{{m_t}^2}}\,\, , \label{cal1}
\enq
\beq
\epsilon_b=\frac{G_F}{2\sqrt{2}{\pi}^2}{m_t}^2
\left ( -\frac{1}{4}\krn+\kln \right ) \log{\frac{{\Lambda}^2}{{m_t}^2}}
\,\, . \label{cal2}
\enq
Note that $\epsilon_2$ and $\epsilon_3$ do not contribute at this order.
That $\klc$ does not contribute to $\epsilon_b$ up to this order
can be understood from Eq.~(\ref{eq3}). If $\klc=-1$
then there is no net \tbw coupling in the chiral lagrangian
after including both the standard and non--standard contributions. Hence,
no dependence on the top quark mass can be generated, {\it i.e.},
the non--standard $\klc$ contribution to $\epsilon_b$
must cancel the SM contribution when
$\klc=-1$, independently of the couplings of
the neutral current.
{}From this observation and because the SM contribution to
$\epsilon_b$ is finite,
we conclude that $\klc$ can not contribute to $\epsilon_b$ at the order
of interest.

Note that we set the renormalization scale $\mu$ to be $m_t$, which is the
natural scale to be used in our study because the top quark is
considered to be the heaviest
mass scale in the effective lagrangian.
We have assumed that all other
heavy fields have been integrated out
to modify the effective couplings of the top quark to gauge bosons
at the scale $m_t$ in the chiral lagrangian.
Here we ignore the effect of
the running couplings from the top quark
mass scale down to the $Z$ boson mass scale which is a
reasonable approximation for our study.

To constrain these non--standard couplings we need to have both the
experimental values and the SM predictions of {\mbox {$\epsilon$'s}}.
First, we tabulate the numerical inputs,
taken from Ref.~\cite{bar}, used in our analysis.
\begin{eqnarray}
{\alpha}^{-1}({M_Z}^2) &=&   128.87 \nonumber\, ,\\
G_F          & =  & 1.166372 \times {10}^{-5} \,\,\,\,
 {\rm{GeV}}^{-2} \nonumber\, ,\\
M_Z          & =  &  91.187 \pm 0.007  \,\,\, \rm{GeV} \nonumber \, ,\\
M_W/M_Z      & =  &  0.8798 \pm 0.0028   \nonumber \, ,\\
\Gamma_{\ell}& =  &  83.52 \pm 0.28  \,\,\, \rm{MeV} \nonumber \, ,\\
\Gamma_b     & =  &  383 \pm 6  \,\,\,\rm{MeV}   \nonumber \, ,\\
{A^{\ell}}_{FB}       & =  &  0.0164 \pm 0.0021 \nonumber \, ,\\
{A^{b}}_{FB}       & =  &  0.098 \pm 0.009  \nonumber \, .
\end{eqnarray}
{}From these values we have \cite{bar}
\begin{eqnarray}
\epsilon_1 \,10^{3}  & =  &  -0.3 \pm 3.4 \nonumber \, ,\\
\epsilon_b \,10^{3}  & =  &  4.4 \pm 7.0 \nonumber \, ,
\end{eqnarray}
and for completeness
\begin{eqnarray}
\epsilon_2 \,10^{3}  & =  &  -7.6 \pm 7.6 \nonumber \, ,\\
\epsilon_3\, 10^{3}  & =  &  0.4 \pm 4.2 \nonumber \, .
\end{eqnarray}

The SM contribution to $\epsilon$'s have been calculated in
Refs.~\cite{bar,bar1}. We will include these contributions in
our analysis in accordance with the assumed Higgs boson mass.
In the light Higgs boson case ($m_H < m_t$), the
calculated values of the $\epsilon$'s include both the SM contribution
calculated in Refs.~\cite{bar,bar1}
and the new physics contribution derived from
the effective couplings of the top quark to gauge bosons.
In the heavy Higgs boson case
($m_H > m_t$) we subtract
the Higgs boson contribution from the SM
calculations of $\epsilon$'s given in {\mbox {Refs.~\cite{bar,bar1}}}.
In this case, the Higgs  boson
contribution is implicitly included in the effective couplings of the top
quark to gauge bosons after the heavy Higgs boson field is integrated
out.
Finally, in a spontaneous symmetry scenario without a Higgs boson
the calculations of {\mbox {$\epsilon$'s}} are exactly the same as those done
in the
heavy Higgs boson case except that the effective couplings of the top quark
to gauge bosons are not due to an assumed heavy Higgs boson in the full
theory.

Choosing $m_t=150$\,GeV and $m_H=100$\,GeV
we span the parameter space  defined by $-1 \leq \kln \leq 1 $,
$-1 \leq \krn \leq 1 $ and $-1 \leq \klc \leq 1 $.
Within $95$\%~C.L. and including both the SM and the new physics
contributions, the allowed region of these three parameters is found
to form a thin slice in the specified volume.
The two--dimensional projections of this slice are shown
in Figs.~(\ref{pr1}),
(\ref{pr2}) and (\ref{pr3}). These non--standard
couplings {\mbox {($\kappa$'s)}} do exhibit
some interesting features:
\begin{itemize}
\item [1)]
As a function of the top quark mass,
the allowed volume for the top quark
couplings to gauge bosons shrinks as the top quark
becomes more massive.
\item [2)]
 New physics prefers positive $\kln$,
see Figs.~(\ref{pr1}) and (\ref{pr2}).
$\kln$ is constrained within $-0.3$ to 0.6 ($-0.2$ to 0.5) for
a 150 (175)\,GeV top quark.
\item [3)]
New physics prefers $\klc \approx -\krn$.
This is clearly shown in Fig.~(\ref{pr3}) which is
the projection of the allowed volume
in the plane containing $\krn$ and $\klc$.
\end{itemize}

In Ref.~\cite{pczh}, a similar analysis has been
carried out by Peccei {\it et al}. However, in their analysis they did
not include the charged current contribution and
assumed only the vertex
\ttz gives large non--standard effects.
The allowed region they found  simply
corresponds, in our analysis, to the region defined by the intersection
of the allowed volume and the plane
$\klc = 0$. This gives a small area confined in the vicinity of the line
$\kln = \krn$. This can be understood from the
expression of $\epsilon_1$ derived
in Eq.~(\ref{cal1}).
After setting $\klc=0$, we find
\beq
\epsilon_1 \propto \left (\krn - \kln \right )\, .
\enq
In this case we note that the length of
the allowed area
is merely determined by the contribution from
$\epsilon_b$. We will
elaborate on a more quantitative comparison in the second part of this
section.

\subsection{Special case}

The allowed region in the parameter space obtained in Figs.~(2)-(4)
contains all possible new physics (to the order ${m_t}^{2}\log {\Lambda}^2$ )
which can modify the couplings of the top quark to gauge bosons
as described by $\kln$, $\krn$ and $\klc$.
In this section we would like to examine a special class of models
in which an approximate custodial symmetry is assumed
as suggested by low energy data.

 The SM has an additional (accidental) symmetry called the
custodial symmetry which is responsible for the tree level relation
\beq
\rho=\frac{M_W}{M_Z\cos \theta_W}=1\,\,. \label{rho}
\enq
This symmetry is slightly broken at the quantum level by the $SU(2)$ doublet
fermion mass splitting and the hypercharge
coupling $g^{\prime}$ \cite{velt}. Writing $\rho=1 +\delta \rho$,
$\delta\rho$ would vanish to all orders if this symmetry is exact.
 Because low energy data indicates that
$\delta\rho$ is very close to zero
we shall therefore assume an underlying theory with a custodial
symmetry.
In other words we require the global
group $SU(2)_V$ associated with the
custodial symmetry to be a subgroup of the full group characterizing
the full theory. We will assume that the custodial symmetry is broken
by the same factors
which break it in the SM, {\it i.e.}, by the fermion
mass splitting and the hypercharge coupling $g^{\prime}$.

In the chiral lagrangian this assumption of a custodial
symmetry sets  $v_3=v$,
and forces the couplings of the top quark to
gauge bosons ${W_{\mu}}^{a}$
to be equal after turning off the hypercharge and
assuming $m_b = m_t$.
If the dynamics of the symmetry breaking is such
that the masses of the two $SU(2)$ partners $t$ and $b$ remain degenerate
then  we expect new physics to contribute
to the couplings of \ttz and \tbw by the same amount.
However, in reality, $m_b \ll m_t$,
thus the custodial symmetry has to be broken.
We will discuss how this symmetry is broken shortly.
Since we are mainly interested in the
{\it leading} contribution enhanced
by the top quark mass at
the order ${m_t}^{2}\log {\Lambda}^2$, turning the hypercharge
coupling on and off will not affect the final result up to this order.

We can construct the two hermitian operators
$J_{L}$ and $J_{R}$, which transform under G as
\beq
{J_{L}}^{\mu}=-i\Sigma {D_{\mu}\Sigma}^{\dagger}
 \rightarrow g_{L}{J_{L}}^{\mu}{g_{L}}^{\dagger}\,\, ,
\enq
\beq
{J_{R}}^{\mu}=i{\Sigma}^{\dagger} D_{\mu}\Sigma
 \rightarrow g_{R}{J_{R}}^{\mu}{g_{R}}^{\dagger}\,\, ,
\enq
where $g_{L}=e^{i\alpha^{a}\frac{\tau^{a}}{2}} \in SU(2)_{L} $\,and
$g_{R}=e^{iy\frac{\tau^3}{2}}$ (note that $v_3=v$ in $\Sigma$).
In fact, using either $J_{L}$ or $J_R$ will lead to the same result.
Hence, from now on we will only consider $J_R$.
The SM lagrangian can be derived from
\begin{eqnarray}
{\cal L}_{0}=\overline{\Psi_{L}}i\gamma^{\mu}{D_{\mu}}^{L}\Psi_{L}
            +\overline{\Psi_{R}}i\gamma^{\mu}{D_{\mu}}^{R}\Psi_{R}
           -(\overline{\Psi_{L}}\Sigma M\Psi_{R}+h.c) \nonumber \\
-\frac{1}{4}{W_{\mu \nu}}^{a}{W^{\mu \nu}}^{a}-
 \frac{1}{4}B_{\mu \nu}B^{\mu \nu}
  +\frac{v^{2}}{4}{\rm{Tr}}({J_{R}}^{\mu}{J_{R}}_{\mu})\,\, ,
\label{yeq48}
\end{eqnarray}
where $M$ is a diagonal mass matrix. We have chosen
 the left--handed fermion fields to be
the ones defined in Eq.~(\ref{psi})
\beq
\Psi_{L}\equiv \Sigma{t\choose b}_{L}\, .
\enq
The right--handed fermion fields $t_{R}$ and $b_{R}$ coincide with the
original right--handed fields (see Eq.~(\ref{psr})). Also
\beq
{D_{\mu}}^{L}=\partial_{\mu}-ig{W_{\mu}}^{a}\frac{\tau^{a}}{2}
  -i{g}^{\prime}B_{\mu}\frac{Y}{2}\,\, ,
\enq
\beq
{D_{\mu}}^{R}=\partial_{\mu}-i{g}^{\prime}B_{\mu}
\left ( \frac{Y}{2} + \frac{\tau^3}{2} \right )\,\, .
\enq
Note that in the non--linear realized effective theories
using either set of fields ( $\Psi_{L,R}$ or $F_{L,R}$)
to construct a chiral lagrangian will lead to the
same S matrix \cite{cole}.

The lagrangian ${\cal L}_0$ in Eq.~(\ref{yeq48})
is not the most general lagrangian one can construct
based solely on the symmetry of $G/H$. Taking
advantage of the chiral lagrangian approach we can
derive additional interaction terms which
deviate from the SM. This is so because in this formalism
the $SU(2)_L\times U(1)_Y$ symmetry is non--linearly realized
and  only the
$U(1)_{em}$ is linearly realized.

Because the SM is so successful one can think of the SM (without
the top quark)
as being the leading term in the expansion of the effective lagrangian.
Any possible deviation associated with
the light fields can only come through
higher dimensional operators in the lagrangian.
However, this assumption is neither
necessary nor preferable when dealing with the top quark because no
precise data are available to lead to such a conclusion. In this
work we will include non--standard dimension--four
operators for the couplings of the top quark to gauge bosons.
In fact this is all
we will deal with and  we will not consider operators with dimension
higher than four. Note that higher dimensional operators are
naturally suppressed by powers of $1/\Lambda$.

One can write $J_{R}$ as
\beq
{J_{R}}^{\mu}= {{J_{R}}^{\mu}}^{a}\frac{{\tau}^{a}}{2} \label{tra} \, ,
\enq
with
\beq
{{J_{R}}^{\mu}}^{a}={\rm{Tr}}\left ( {\tau}^{a}{J_{R}}^{\mu}\right )
=i{\rm{Tr}}\left ( \tau^{a}{\Sigma}^{\dagger}{\rm{D}}^{\mu}\Sigma \right )\, .
\enq
The full operator $J_{R}$ posses an explicit custodial symmetry
when $g^{\prime}=0$ as can easily be checked by expanding it in powers of
the Goldstone boson fields.

Consider first the left--handed sector. One can add
additional interaction terms to the lagrangian
${\cal L}_0$
\beq
{\cal L}_1 ={\kappa}_{1}\overline{\Psi_{L}}\gamma_{\mu}\Sigma {J_{R}}^{\mu}
         {\Sigma}^{\dagger}\Psi_{L}
     +{\kappa}_{2}\overline{\Psi_{L}}\gamma_{\mu}\Sigma
     {\tau}^{3}{J_{R}}^{\mu}{\Sigma}^{\dagger}\Psi_{L}
     +{\kappa}_{2}^\dagger \overline{\Psi_{L}}
       \gamma_{\mu}\Sigma {J_{R}}^{\mu}{\tau}^{3}
     {\Sigma}^{\dagger}\Psi_{L}
\label{ehab}\, ,
\enq
where ${\kappa}_{1}$ is an arbitrary real parameter and $\kappa_2$ is
an arbitrary complex parameter.
Here we do not include interaction terms like
\beq
 {\kappa}_{3}\overline{\Psi_{L}}\gamma_{\mu}\Sigma {\tau}^{3}{J_{R}}^{\mu}
         {\tau}^{3}{\Sigma}^{\dagger}\Psi_{L}\, ,
\enq
where ${\kappa}_{3}$ is real, because
it is simply a linear combination of the other two terms in ${\cal L}_1$.
This can be easily checked by
using Eq.~(\ref{tra}) and the commutation relations of the Pauli matrices.
Note that ${\cal L}_{1}$ still is not the most general lagrangian one can
write for the left--handed sector, as compared to Eq.~(\ref{eq2}).
In fact, it is our insistence on using
the fermion doublet form and the full operator $J_{R}$ that lead us to this
form. For shorthand, ${\cal L}_{1}$ can be further rewritten as
\beq
{\cal L}_1 =\overline{\Psi_{L}}\gamma_{\mu}\Sigma K_{L}{J_{R}}^{\mu}
         {\Sigma}^{\dagger}\Psi_{L}
         +\overline{\Psi_{L}}\gamma_{\mu}\Sigma {J_{R}}^{\mu} K_{L}^\dagger
         {\Sigma}^{\dagger}\Psi_{L}
\label{cp} \,\, ,
\enq
where $ K_{L}$ is a complex diagonal matrix.

These new terms can be generated either through some
electroweak symmetry breaking scenario or
through some other new heavy physics effects.
If $m_b=m_t$ and $g^{\prime}=0$, then we
require the effective lagrangian to respect fully the custodial
symmetry to all orders. In this limit,
${\kappa}_{2}=0$ in Eq.~(\ref{ehab}) and
{\mbox {$K_{L} =\kappa_{1}${\bf 1}}}, where {\bf 1} is
the unit matrix and $\kappa_1$ is real.

Since $m_b \ll m_t$, we can think of $\kappa_{2}$
as generated through the evolution from $m_b=m_t$ to
$m_b=0$. In the
matrix notation this implies $K_{L}$ is not proportional
to the unit matrix and can be parameterized by
\beq
 K_{L}=\pmatrix{{\kappa_{L}}^{t} & 0 \cr 0 & {\kappa_{L}}^{b}\cr}\, ,
\enq
with
\beq
{\kappa_{L}}^{t}=\frac{\kappa_1}{2} + \kappa_2\, ,
\enq
and
\beq
{\kappa_{L}}^{b}=\frac{\kappa_1}{2} - \kappa_2\, \label{me}.
\enq
In the unitary gauge we get the following terms
\begin{eqnarray}
+\frac{g}{2c}2{\rm{Re}} ({\kappa_{L}}^{t})\overline{t_{L}}
\gamma^{\mu}t_{L}Z_{\mu}
+\frac{g}{\sqrt{2}}({\kappa_{L}}^t+{{\kappa_{L}}^{b}}^{\dagger})
\overline{t_{L}}\gamma^{\mu}b_{L}{W_{\mu}}^{+} \nonumber \\
+\frac{g}{\sqrt{2}}({\kappa_{L}}^{b}+{{\kappa_{L}}^{t}}^{\dagger})
\overline{b_{L}}\gamma^{\mu}t_{L}{W_{\mu}}^{-}
-\frac{g}{2c}2 {\rm{Re}} ({\kappa_{L}}^{b})
\overline{b_{L}}\gamma^{\mu}b_{L}Z_{\mu}\, .
\end{eqnarray}
As discussed in the previous section,
we will assume that new physics effects will not modify
the $b_{L}$-$b_{L}$-$Z$ vertex. This can be achieved by
choosing $\kappa_1=2{\rm{Re}}(\kappa_2)$ such that
${\rm{Re}}({\kappa_{L}}^{b})$ vanishes
in Eq.~(\ref{me}). Later, in
{\mbox {section IV}}, we will consider a specific model to support this
assumption.

 Since the imaginary parts of the couplings do not contribute to LEP
physics of interest, we simply drop them hereafter.
With this assumption we are left with
one real parameter ${\kappa_{L}}^{t}$ which will be
denoted from now on by $\kappa_L/2$. The left--handed top quark couplings to
the gauge bosons are
\beq
 t_{L}-t_{L}-Z :\,\, \frac{g}{4c}\kappa_{L}\gamma_{\mu}(1-\gamma_{5})\, ,
\enq
\beq
 t_{L}-b_{L}-W :\,\, \frac{g}{2\sqrt{2}}\frac{\kappa_{L}}{2}
 \gamma_{\mu}(1-\gamma_{5})\, .
\enq
Notice the connection between the neutral and the
charged current, as compared to Eq.~(\ref{eq3}),
\beq
\kln = 2\klc = \kappa_{L}\, .
\enq
This conclusion holds for any underlying theory with an approximate
custodial symmetry such that the vertex \bbz is not modified as discussed
above.

For the right--handed sector, the situation is different because the
right--handed fermion
fields are $SU(2)$ singlet, hence the induced interactions do not see
the full operator $J_{R}$ but its components individually.
Therefore, we cannot impose the
previous connection between the neutral
and charged current couplings.

The additional allowed interaction terms in
the right--handed sector are given by
\begin{eqnarray}
{\cal L}_{2}=
\frac{g}{2c}{{\kappa_{R}}^t}^{NC} \overline{t_{R}}\gamma^{\mu}
t_{R} {{J_{R}}_{\mu}}^{3}
+\frac{g}{\sqrt{2}}{\kappa_{R}}^{CC}\overline{t_{R}}\gamma^{\mu}b_{R}
{{J_{R}}_{\mu}}^{+} \nonumber \\
+\frac{g}{\sqrt{2}}{{\kappa_{R}}^{CC}}^{\dagger}\overline{b_{R}}\gamma^{\mu}
t_{R}{{J_{R}}_{\mu}}^{-}
-\frac{g}{2c} {{\kappa_{R}}^b}^{NC} \overline{b_{R}}\gamma^{\mu}
b_{R} {{J_{R}}_{\mu}}^{3} \label{neh}\,\, ,
\end{eqnarray}
where ${{\kappa_{R}}^{t}}^{NC}$
and  ${{\kappa_{R}}^{b}}^{NC}$ are two arbitrary
real parameters and ${\kappa_{R}}^{CC}$ is an arbitrary complex parameter.
Note that in ${\cal L}_{2}$ we have one more additional coefficient than
we have in ${\cal L}_{1}$ (in Eq.~(\ref{ehab})),
this is due to our previous assumption of
using the full operator $J_{R}$ in constructing the left--handed interactions.
We assume that the \brbrz vertex just as the \blblz vertex
is not modified, then
the coefficient ${{\kappa_{R}}^b}^{NC}$ vanishes.
Because $\krc$ does not contribute to LEP physics in the limit of
$m_b=0$ and at the order ${m_t}^2 \log {\Lambda}^2$ we are left with
one real parameter ${{\kappa_{R}}^t}^{NC}$ which will be
denoted hereafter as $\kappa_{R}$.
The right--handed top quark coupling to $Z$ boson is
\beq
t_{R}-t_{R}-Z :\,\,
\frac{g}{4c}  \kappa_{R}\gamma_{\mu}(1+\gamma_{5}) \,\, .
\enq

In the rest of this section, we consider models described by
${\cal L}_{1}$ and ${\cal L}_{2}$ with only two relevant parameters
$\kappa_L$ and $\kappa_R$. Performing the calculations as we discussed in
the previous subsection we find
\beq
\epsilon_{1} = \frac{G_F}{2\sqrt{2}{\pi}^2} 3 {m_{t}}^2
 (\kappa_{R} - \frac{\kappa_{L}}{2})\log(\frac{{\Lambda}^2}{{m_{t}}^2})\,\, ,
\label{fn}
\enq
\beq
\epsilon_{b} = \frac{G_F}{2\sqrt{2}{\pi}^2}{m_{t}}^2
 (-\frac{1}{4} \kappa_{R} + \kappa_{L})\log(\frac{{\Lambda}^2}{{m_{t}}^2})
 \label{sb} \,\, .
\enq
These results simply correspond to those
in Eqs.~(\ref{cal1}) and (\ref{cal2})
after substituting $\kln=2\klc=\kappa_{L}$ and $\krn=\kappa_R$.

The constraints on $\kappa_L$ and $\kappa_R$
for models with a light Higgs boson
or a heavy Higgs boson,
or without a physical scalar (such as a
Higgs boson) are presented here in order.
Let us first consider a standard light Higgs boson
with mass $m_H=100$\,GeV. Including
the SM contribution from Ref.~\cite{bar} we span the plane
defined by $\kappa_{L}$ and $\kappa_{R}$ for top mass 150 and 175\,GeV,
respectively.
Figs.~(\ref{f5}) and (\ref{f6}) show the allowed range for those parameters
within 95\% C.L.
As a general feature one observes that the allowed range is a
narrow area aligned close to the line $\kappa_{L} = 2 \kappa_{R}$ where for
$m_t=150$\,GeV the maximum  range
for $\kappa_{L}$ is between $-0.1 $ and $0.5$. As the
top mass increases this range shrinks and moves downward and to the
right away form the origin ($\kappa_{L}$,$\kappa_{R}$) = {\mbox {(0,0)}}.
The deviation from the relation $\kappa_{L} = 2 \kappa_{R}$ for
various top quark masses is given in Fig.~(\ref{f7}) by calculating
$\kappa_{L} - 2 \kappa_{R}$ as a function of $m_t$. Note that the
SM has the solution $\kappa_{L}=\kappa_R=0$, {\it i.e.}, the SM solution
lies on the horizontal line shown in Fig.~(7). This solution
ceases to exist for $m_{t}\geq 200$\,GeV.
The special relation $\kappa_{L}=2\kappa_{R}$ is
a consequence of the assumption we imposed in connecting the left--handed
neutral and charged current. The range of the allowed couplings
is summarized in Table~1 and for different top mass.

It is worth mentioning
that the SM contribution to $\epsilon_{b}$ is lower
than the experimental central value \cite{bar,bar1}.
This is reflected in the behavior of
$\kappa_{L}$ which prefers being positive to compensate this difference as
can be seen from Eq.~(\ref{sb}).
This means in models of electroweak symmetry breaking with an approximate
custodial symmetry, a positive $\kappa_L$ is preferred.
In Fig.~(\ref{f8}) we show the allowed $\klc=\kln/2=\kappa_{L}/2$ as a function
of $m_t$.
With new physics effects ($\kappa_L \neq 0$) $m_t$ can be as large as
300\,GeV, although in the SM ($\kappa_L=0$) $m_t$ is bounded below
200\,GeV.

Now, we would like to discuss the effect of a light SM Higgs boson
($ m_H < m_t$) on the allowed range of these parameters.
It is easy to anticipate the effect; since
$\epsilon_{b}$ is not sensitive to the Higgs contribution up to one loop
\cite{bar},
the allowed range is only affected by the Higgs contribution to
$\epsilon_{1}$ which affects slightly the
width of the allowed area and its location relative to the line
$\kappa_{L}=2\kappa_{R}$.
One expects that as the Higgs mass increases the allowed area moves
upward. The reason simply lies in the fact that the standard Higgs boson
contribution to $\epsilon_{1}$ up to one loop becomes more negative
for heavier Higgs boson,
hence $2\kappa_{R}$ prefers to be larger than $\kappa_{L}$
to compensate this effect. However, this
modification is not significant because $\epsilon_{1}$
depends on the
Higgs boson mass only logarithmically \cite{bar1}.

If there is a heavy
Higgs boson ($m_H > m_t$), then it should be integrated out
from the full theory  and its effect in the chiral lagrangian is
manifested through the effective couplings of the top quark to gauge
bosons.
In this case we simply
subtract the Higgs boson contribution from the
SM results obtained in Refs.~\cite{bar,bar1}.
Fig.~(\ref{f9}) shows the allowed area in the $\kappa_L$ and $\kappa_R$ plane
for a 175\,GeV top quark in such models.
Again we find no noticeable difference between the results from these models
and those with a light Higgs boson. That is because
up to one loop level neither $\epsilon_1$
nor $\epsilon_b$ is sensitive
to the Higgs boson contribution \cite{bar,bar1}.

If we consider a new
symmetry breaking scenario without a fundamental scalar such
as a SM Higgs boson,
following the previous discussions we again find
negligible effects on the allowed range of $\kappa_L$ and $\kappa_R$.

What we learned is that
to infer a bound on the Higgs boson mass
from the measurement of the effective couplings of the top quark to gauge
bosons, we need very precise measurement of the parameters $\kappa_L$
and $\kappa_R$.
However, from the correlations between the effective
couplings {\mbox {($\kappa$'s)}}
of the top quark to gauge bosons, we can infer if
the symmetry breaking sector is due to a Higgs boson or not, {\it i.e.},
we may be able
to probe the symmetry breaking mechanism in the top quark system.
Further discussion will be given in the next section.

Finally, we would like to compare our results with those in
Ref.~\cite{pczh}. Fig.~(\ref{f10}) shows the most general allowed region for
the couplings $\kln$ and $\krn$, {\it i.e.}, without imposing any relation
between $\kln$ and $\klc$. This region is for top mass 150\,GeV and is
covering the parameter space $-1.0 \leq \kln \, , \krn \leq 1.0 $. We find
\begin{eqnarray}
-0.3 &\leq& \kln \leq 0.6 \, ,\nonumber \\
-1.0 &\leq& \krn \leq 1.0 \nonumber \, .
\end{eqnarray}
Also shown on Fig.~(10) the allowed regions from our model and
the model in Ref.~\cite{pczh}.
The two regions overlap in the vicinity of the origin (0, 0) which
corresponds to the SM case.
As $\kln \geq 0.1$, these two regions diverge and become separable.
One notices that the allowed range predicted in Ref.~\cite{pczh} lies
along the line $\kln=\krn$ whereas in our case the slope is different
$\kln =2\krn$. This difference comes in because of the assumed dependence
of $\klc$ on the other two couplings $\kln$ and $\krn$. In our case
$\klc =\kln/2$, and in Ref.~\cite{pczh} $\klc=0$.

Note that for $m_{t}\leq 200$\,GeV the allowed region of
{\mbox {$\kappa$'s}}
in all models of symmetry breaking should overlap
near the origin because the SM is consistent with low energy data at
the 95\% C.L. If we imagine that any prescribed dependence between the
couplings corresponds to a symmetry breaking scenario, then, given the
present status of low energy data, it is
possible to distinguish between different scenarios if $\kln$, $\krn$ and
$\klc$ are larger than 10\%. Better future measurements of
{\mbox {$\epsilon$'s}} can further discriminate between different
symmetry breaking scenarios. We will discuss how the SLC,
the Tevatron, the LHC and the NLC can contribute to
these measurements in section V.
Before that, let us examine a specific model that predicts certain relations
among the coefficients $\klc$, $\krc$, $\kln$ and $\krn$
of the effective couplings of the top quark to gauge bosons.

\section{ Heavy higgs limit in the SM }

The goal of this study
is to probe new physics effects, particularly the effects due
to the symmetry breaking sector, in the top quark system by examining
the couplings of top quark to gauge bosons.
To illustrate how a specific symmetry breaking mechanism might affect
these couplings, we consider in this section the Standard Model with a
heavy Higgs boson ($m_H > m_t$) as the full theory,
and derive the effective couplings
$\kln$, $\krn$, $\klc$ and $\krc$ at the top quark mass scale in the effective
lagrangian after integrating out the heavy Higgs boson field.

Given the full theory (SM in this case), we can perform matching between
the underlying theory and the effective lagrangian.
In this case, the heavy Higgs boson mass
acts as a regulator (cut--off) of the effective theory\cite{long}.

While setting $m_{b}=0 $, and
only keeping the leading terms of the order
${m_t}^{2}\log{m_H}^{2}$,
we find the following effective couplings
\beq
 t-t-Z:\,\,
\frac{g}{4c}\frac{G_F}{2\sqrt{2}{\pi}^2}\left ( \frac{-1}{8}{m_t}^{2}
 \gamma_{\mu}(1-\gamma_{5})
  +\frac{1}{8}{m_t}^{2}\gamma_{\mu}(1+\gamma_{5}) \right )
  \log \left(\frac{{m_H}^{2}}{{m_t}^{2}}\right)\, ,
\enq
\beq
 t-b-W:\,\,\frac{g}{2\sqrt{2}}
\frac{G_F}{2\sqrt{2}{\pi}^2}\left(\frac{-1}{16}\right){m_t}^2
\gamma_{\mu}(1-\gamma_{5})\log\left(\frac{{m_H}^{2}}{{m_t}^{2}}\right)\, .
\enq
{}From this we conclude
\beq
{\kappa_{L}}^{NC}=2{\kappa_{L}}^{CC}= \frac {G_F}{2\sqrt{2}{\pi}^2}
 \left(\frac{-1}{8}\right){m_t}^2
 \log\left(\frac{{m_H}^{2}}{{m_t}^{2}}\right)\, ,
\enq
\beq
{\kappa_{R}}^{NC}=\frac{G_F}{2\sqrt{2}{\pi}^2}
\frac{1}{8}{m_t}^2
 \log\left(\frac{{m_H}^{2}}{{m_t}^{2}}\right)\, ,
\enq
\beq
{\kappa_{R}}^{CC}=0\, .
\enq

 Note that the relation between the left--handed currents
($\kln=2\klc$) agree with our
prediction because of the approximate custodial symmetry
in the full theory (SM) and the fact that vertex \bbz is not modified. The
right--handed currents $\krc$ and $\krn$ are not correlated,
and $\krc$ vanishes for a massless $b$.
Also note an additional relation in the effective lagrangian
between the left-- and right--handed effective
couplings of the top quark to $Z$ boson, {\it i.e.},
\beq
\kln =-\krn \, .
\enq
This means only the axial vector current of \ttz acquires a non--universal
contribution but its vector current is not modified.

As discussed in section  II, due to the Ward identities
associated with the photon field
there can be no non--universal contribution to either
the \bba or \tta vertex
after renormalizing the fine structure constant $\alpha$.
This can be explicitly checked in this model.
Furthermore, up to the order of ${m_t}^{2}\log{m_H}^{2}$, the
vertex \bbz is not modified which agrees with the assumption we made in
section II that there exist dynamics of electroweak symmetry breaking so
that neither \brbrz nor \blblz in the effective lagrangian
is modified at the scale of $m_t$.

{}From this example we learn that the effective couplings of the
top quark to gauge bosons arising from
a heavy Higgs boson are correlated in a specific way, namely
\beq
\kln=2 \klc=-\krn \, .
\enq
(This relation in general
also holds for models with a heavy scalar which is not necessarily a SM
Higgs boson, {\it i.e.}, the coefficients of the last two terms in
Eq.~(\ref{higg})
can be arbitrary, and are not necessarily $1/2$ and $1/4$, respectively).
In other words, if the couplings of a heavy
top quark to the gauge bosons are measured
and exhibit large deviations from these
relations, then it is likely that the electroweak
symmetry breaking is not due to the standard Higgs mechanism
which contains a heavy SM Higgs boson.
This illustrates how the symmetry breaking sector can be probed
by measuring the effective couplings of the top quark to gauge bosons.

\section{ Direct Measurement of the Top Quark Couplings}

In section III we concluded that the precision LEP
data can constrain the couplings $\kln$, $\krn$ and $\klc$,
but not $\krc$ (the right--handed charged current).
In this section, we examine how to
improve our knowledge on these couplings at the current and future
colliders.

\subsection{ At the SLC }

The measurement of the left--right cross section asymmetry $A_{LR}$
in $Z$ production with a longitudinally polarized electron beam
at the SLC provides a stringent test of the SM and is sensitive to new
physics.

Additional constraints on the couplings $\kln$, $\krn$
and $\klc$ can be inferred from  $A_{LR}$ which can be written as
\cite{bar}
\beq
A_{LR} = \frac{2x}{1+x^2} \, ,
\enq
with
\beq
x=1-4s^2(1+\Delta k^{\prime} ) \,\, ,
\enq
\beq
\Delta k^{\prime} = \frac{\epsilon_{3}- c^2\epsilon_{1}}{c^2 - s^2} \, .
\enq
Up to the order ${m_t}^{2}\log {\Lambda}^{2}$, only
$\epsilon_1$ contributes. In our model with the approximate custodial
symmetry, {\it i.e.}, $\kln= 2 \klc=\kappa_L$,
the SLC $A_{LR}$ measurement will have a significant influence on the
precise measurement of the non--universal couplings of the top quark.
This influence
will be through decreasing the width of the allowed area in the
parameter space ($\kappa_L$ versus $\kappa_R$)
shown in Figs.~(\ref{f5}) and (\ref{f6}).
For instance, with an expected uncertainty 0.001
in the 1993 run on the measurement of the
effective electroweak mixing angle,
$\sin^2 \theta_w^{eff} =(1-x)/4 $, at the SLC \cite{thesis},
the width of the allowed area shown in Figs~.(\ref{f5}) and (\ref{f6})
will be shrunk by more than a factor of 5.
However, there will be no effect on
the length of the allowed region which in our approximation is solely
determined by $\epsilon_b$.
Hence, a more accurate measurement of $\epsilon_b$, {\it i.e.},
$\Gamma (Z \rightarrow b \bar b) $,
is required to further confine the non--universal interactions of
the top quark to gauge bosons to probe new physics.
%%%%%%%%%%%%%%%%%%%%%%%%%%%%%%%%%%%%%%%%%%%

\subsection{ At the Tevatron and the LHC}

In this section, we study how to constrain the
non--standard couplings of the top quark to gauge bosons
from direct detection of the top quark at hadron colliders.

At the Tevatron and the LHC, heavy top quarks are
predominantly produced from the QCD process
$gg, q \bar q \ra t \bar t$ and the $W$--gluon fusion process
$qg (Wg) \ra t \bar{b}, \bar{t} b$.
In the former process, one can probe $\klc$ and $\krc$ from the decay of the
top quark to a bottom quark and a $W$ boson. In the latter process,
these non--standard couplings can be measured by simply counting the production
rates of signal events with a single $t$ or $\bar t$.
More details can be found in Ref.~\cite{anlrev}.

To probe $\klc$ and $\krc$ from the decay of the
top quark to a bottom quark and a $W$ boson, one needs to measure the
polarization of the $W$ boson. For a massless $b$, the $W$ boson from top
quark decay can only be either longitudinally or left--handed polarized for
a left--handed charged current ($\krc=0$). For a right--handed
charged current ($\klc=-1$) the $W$ boson can only be either longitudinally
or right--handed polarized.
(Note that the handedness of the $W$ boson is reversed for a massless
$\bar b$ from $\bar t$ decays.)
In all cases the fraction of longitudinal $W$
from top quark decay is enhanced by ${m_t}^2/{2{M_W}^2}$ as compared
to the fraction of transversely polarized $W$. Therefore, for
a more massive
top quark, it is more difficult to untangle the $\klc$ and $\krc$
contributions. The $W$ polarization measurement can be done by measuring the
invariant mass ($m_{b\ell}$)
of the bottom quark
and the charged lepton from the decay of top quark \cite{kanetop}.
We note that this method does not require knowing the longitudinal momentum
(with two--fold ambiguity)
of the neutrino from $W$ decay to reconstruct the rest frame of the
$W$ boson in the rest frame of the top quark.

Consider the (upgraded)
Tevatron as a $\ppbar$ collider at
$\sqrt{S}= 2$ or 3.5 TeV,
 with an integrated luminosity of $1\,{\rm or}\,10\,$
$\rm{fb}^{-1}$.
Unless specified otherwise,
we will give event numbers for a 175 GeV top quark and
an integrated luminosity of 1 $\rm{fb}^{-1}$.

The cross section of the QCD process $gg,q\bar{q}\ra t \bar t$
is about 7 (29) pb at a $\sqrt{S}= 2$ (3.5) TeV collider.
In order to measure $\klc$ and $\krc$ we have to
study the decay kinematics of the
 reconstructed $t$ and/or $\bar t$.
For simplicity, let's consider the $\ell^\pm \, + \geq 3\, {\rm jet}$
decay mode, whose
branching ratio is $Br=2 {\frac{2}{9}} {\frac{6}{9}} = \frac{8}{27}$,
for $\ell^+=e^+ \,{\rm or}\,\mu^+$.
We assume an experimental detection
efficiency, which includes
both the kinematic acceptance and the efficiency of $b$--tagging,
of  15\% for the $t \bar t$ event. We further assume that there is
no ambiguity in picking up the right $b$ ($\bar b$)
to combine with the charged lepton $\ell^+$ ($\ell^-$)
to reconstruct $t$ ($\bar t$). In total, there are
$7\,{\rm pb}\,\times \, 10^3\,{\rm pb}^{-1}\,
\times \,{\frac{8}{27}}\,\times \,0.15=300$
reconstructed $t \bar t$ events to be used in measuring
$\klc$ and $\krc$ at $\sqrt{S}= 2$\,TeV.
The same calculation at
$\sqrt{S}= 3.5$\,TeV yields $1300$ reconstructed $t \bar t$ events.
Given the number of reconstructed top quark events,
one can in principle fit the $m_{b\ell}$ distribution to measure
$\klc$ and $\krc$. We note that the polarization of the $W$ boson can also be
studied from
the distribution of $\cos \theta^*_\ell$,
where $\theta^*_\ell$ is the
polar angle of $\ell$ in the rest frame of the $W$ boson whose z--axis
is the $W$ bosons moving direction in the rest
frame of the top quark \cite{kanetop}.
For a massless $b$,  $\cos \theta^*_\ell$ is related to ${m_{b\ell}}^2$ by
\beq
{\cos \theta^*_{\ell}}
\simeq {{{2 {m_{b\ell}}^2}\over{{m_t}^2-{M_W}^2}} - 1}\, .
\enq

However, in reality, the momenta of the bottom quark and the
charged lepton will be smeared by
the detector effects and the most serious problem in this analysis is
the identification of the right $b$ to reconstruct $t$.
There are two strategies to improve the efficiency of identifying
the right $b$.  One is to demand a large invariant mass of the $t \bar t$
system so that $t$ is boosted and its decay products are collimated.
Namely, the right $b$ will be moving closer to the lepton from $t$ decay.
This can be easily enforced by demanding lepton $\ell$
 with large transverse
momentum.
Another is to identify the non--isolated lepton from $\bar b$ decay
(with a branching ratio $Br(\bar b \ra \mu^{+} X) \sim 10\%$).  Both
of these methods will further reduce the reconstructed signal rate by
an order of magnitude.
How will these affect our conclusion on
the determination of the non--universal couplings $\klc$ and $\krc$?
This cannot be answered in the absence of
detailed Monte Carlo studies.

Here we propose to probe the
couplings $\klc$ and $\krc$ by measuring the
production rate of the single--top quark events.
A single--top quark event
 can be produced from either the $W$--gluon fusion process
$q g \, (W^+g) \ra t \bar{b} X$, or the Drell--Yan type process
$q \bar q \ra W^* \ra t \bar b$.
Including both the single--$t$ and single--${\bar t}$ events,
for a 2 (3.5) TeV collider, the $W$--gluon fusion rate
is 2 (16) pb;
the Drell--Yan type rate is 0.6 (1.5) pb.
The Drell--Yan type event is easily separated from the
$W$-gluon fusion event, therefore will
not be considered hereafter \cite{carlson}.
For the decay mode of
$t \ra b W^+ \ra b \ell^+ \nu$, with $\ell^+=e^+ \,{\rm or}\,\mu^+$,
the branching ratio of interest is
$Br=\frac{2}{9}$.
The kinematic acceptance
of this event at $\sqrt{S}= 2$\,TeV is
found to be $0.55$ \cite{carlson}.
If the efficiency of $b$--tagging is 30\%,
there will be
$2\,{\rm pb}\,\times \,10^3\,{\rm pb}^{-1}\,
\times \,{\frac{2}{9}}\,\times \,0.55\, \times \,0.3=75$
single-top quark
events reconstructed. At $\sqrt{S}= 3.5$\,TeV the kinematic acceptance
of this event is $0.50$ which,
from the above calculation yields
about $530$ reconstructed events.
Based on statistical
error alone, this corresponds to a 12\% and 4\%
measurement on the single--top cross section.
A factor of 10 increase in the luminosity of
the collider can improve the measurement by a factor of 3 statistically.

Taking into account the theoretical uncertainties, we
examine two scenarios: 20\% and 50\% error on
the measurement of the single--top cross section,
which depends on both $\klc$ and $\krc$.
(Here we assume the experimental data
agrees with the SM prediction within 20\% (50\%).)
We found that for a $175\,$GeV top quark
$\klc$ and $\krc$ are well constrained inside
the region bounded by two (approximate) ellipses,
 as shown in Fig.~(4).
These results are not sensitive to
the energies of the colliders considered here.

The top quark produced from the $W$--gluon fusion process
is almost one hundred percent left--handed (right--handed) polarized
for a left--handed (right--handed) $\tbw$ vertex, therefore
the charged lepton $\ell^+$ from $t$ decay has a harder momentum
in a right--handed $\tbw$ coupling than in a left--handed coupling.
(Note that the couplings of
light--fermions to $W$ boson have been well tested
from the low energy data to be left--handed as described in the SM.)
This difference becomes smaller when the top quark is
more massive  because
the $W$ boson from the top quark decay tends to be more
longitudinally polarized.

A right--handed charged current is absent in a
linearly $SU(2)_L$ invariant gauge
theory with massless bottom quark.
In this case,  $\krc=0$,
then $\klc$ can be constrained to within
about $-0.08 < \klc < 0.03$ ($-0.20 < \klc < 0.08$)
with a 20\% (50\%) error on the measurement of the
the single--top quark production rate
 at the Tevatron. This means that if
we interpret {\mbox {($1+\klc$)}} as the CKM matrix element $V_{tb}$,
then $V_{tb}$ can be bounded as $V_{tb} > 0.9$ (or 0.8) for a 20\%
(or 50\%)  error on the measurement of
the single--top production rate.
Recall that if there are more than three generations,
within  90\% C.L.,
$V_{tb}$ can be anywhere between 0 and 0.9995 from low energy data
\cite{book}. This measurement can therefore provide useful information on
 possible additional fermion generations.

We expect the LHC can provide similar or better bounds on these
non--standard couplings when detail analyses are available.
%%%%%%%%%%%%%%%%%%%%%%%%%%%%%%%%%%%%%%%%%%%%%%%%%%%%%%%%%%%%

\section{ At the NLC }

The best place to probe $\kln$ and $\krn$ associated with the
\ttz coupling is at the NLC through $e^- e^+ \ra A, Z \ra t \bar{t}$.
(We use NLC to represent a generic $e^-e^+$ supercollider \cite{nlc}.)
A detail Monte Carlo study on the measurement of these couplings at the NLC
including detector effects and initial state radiation
can be found in Ref.~\cite{gal}.
The bounds were obtained by studying the
angular distribution and the polarization of the top quark
produced in $e^- e^+$ collisions.
Assuming a 50 $\rm{fb}^{-1}$
luminosity at $\sqrt{S}=500$\,GeV, we concluded that
within 90\% confidence level, it should be possible to measure
$\kln$ to within about 8\%, while $\krn$ can be known
to within about 18\%.
A $1\,$TeV machine can do better than a $500\,$GeV machine in
determining $\kln$ and $\krn$ because the relative sizes of the
$t_R {(\overline{t})}_R$  and $t_L {(\overline{t})}_L$
production rates become small and the polarization of the $t \bar t$ pair
is purer. Namely, it's more likely to produce either
a $t_L {(\overline{t})}_R$ or a $t_R {(\overline{t})}_L$ pair.
A purer polarization of the $t \bar t$ pair makes $\kln$ and $\krn$
better determined. (The purity of the $t \bar t$ polarization
can be further improved by polarizing the electron beam.)
Furthermore, the top quark is
boosted more in a $1\,$TeV machine thereby allowing a better
determination of its polar angle in the $t \bar t$ system
because it is easier to find
the right $b$ associated with the lepton to reconstruct the
top quark moving direction.

Finally, we remark that at the NLC, $\klc$ and $\krc$ can be studied
either from the decay of the top quark pair or from the single--top
quark production process; $W$--photon fusion
process $e^{-}e^{+}(W\gamma) \ra t X $,
or $ e^{-}\gamma (W\gamma) \ra {\bar t} X$, which is similar to the
$W$--gluon fusion process in hadron collisions.

\section{discussion and conclusions}

In this paper, we have applied the electroweak chiral lagrangian to probe
new physics beyond the SM through studying the couplings of the top quark
to gauge bosons.
We first examined the precision LEP data to extract the information on these
couplings, then we discussed how to improve our knowledge at current and
future colliders such as at the Tevatron, the LHC and the NLC.

Due to the non--renormalizability of the electroweak chiral lagrangian, we
can only estimate the size of these non--standard couplings by studying the
contributions to LEP observables at the order of
${m_t}^{2}\log{\Lambda}^{2}$, where $\Lambda = 4 \pi v \sim 3$ TeV
is the cut--off scale of the effective lagrangian.
Already we found interesting constraints on these couplings.

Assuming \bbz vertex is not modified, we found that $\kln$ is already
constrained to be $-0.3 < \kln < 0.6$ ($-0.2 < \kln < 0.5$)
by LEP data at the 95\% C.L.
for a 150 (175)\,GeV top quark.
Although $\krn$ and $\klc$ are allowed to be in the full
range of $\pm 1$, the precision LEP data do impose some correlations among
$\kln$, $\krn$ and $\klc$. ($\krc$ does not contribute to the LEP
observables of interest in the limit of $m_b=0$.)
In our calculations, these non--standard couplings are only inserted once
in loop diagrams using dimensional regularization.

Inspired by the experimental fact $\rho \approx 1$, reflecting
the existence of an approximate custodial symmetry, we
proposed an effective model to relate $\kln$ and $\klc$.
We found that the non--universal
interactions of the top quark to
gauge bosons parameterized by $\kln$, $\krn$ and $\klc$
are well constrained by LEP data, within 95\% C.L.
The results are summarized in {\mbox {Table 1}} (see also Figs.~(\ref{f5})
and (\ref{f6})).
Also, the two parameters $\kappa_L=\kln$ and $\kappa_R=\krn$ are strongly
correlated. In our model, $\kappa_L \sim 2 \kappa_R$.

We note that the relations among $\kappa$'s
can be used to test different models of
electroweak symmetry breaking. For instance, a heavy SM Higgs boson
($m_H > m_t$) will
modify the couplings \ttz and \tbw of
a heavy top quark at the scale $m_t$ such that
$\kln = 2\klc$, $\kln =-\krn$ and $\krc=0$.

It is also interesting to note that the upper bound on the top quark mass
can be raised from the SM bound $m_t < 200$\,GeV to as large as
300\,GeV if new physics occurs.
That is to say, if there is new physics
associated with the top quark, it is
possible that the top quark is heavier
than what the SM predicts, a similar
conclusion was reached in Ref.~\cite{pczh}.

With a better measurement of $A_{LR}$ at the SLC,
more constraint can be set on the correlation between
$\kappa_L$ and $\kappa_R$. To constrain the size of
$\kappa_L$ and $\kappa_R$, we need
a more precise measurement on the partial decay
width $\Gamma (Z \ra b \bar b)$.

Undoubtedly, direct detection of the top quark at the Tevatron,
the LHC and the NLC is crucial to measuring the couplings of
\tbw and \ttz. At hadron colliders, $\klc$ and $\krc$ can be measured by
studying the polarization of the $W$ boson from top quark decay
in $t \bar t$ events. They can also be measured simply from the production
rate of the single top quark event.
The NLC is the best machine to measure $\kln$ and $\krn$ which can be
measured from studying the angular distribution and the polarization
of the top quark produced in $e^- e^+$ collision.
Details about these bounds were given in section V.

\section*{ Acknowledgments }

We thank
R. Brock, D.O. Carlson, R.S. Chivukula, M. Einhorn, K. Lane, E. Nardi,
E.H. Simmons, M. Wise and Y.--P. Yao
for helpful discussions. We also thank A. Abbasabadi and W. Repko
for a critical reading of the manuscript.
This work was supported in part by an NSF grant No. PHY-9309902.

%%%%%%%%%%%%%%%%%%%%%%%%%%%%%%%%%%%%%%%%%%%%%%%%%%%%%%%%%%%%%%%%%%%
\begin{figure}
\caption{ Some of the relevant Feynman diagrams in
the 't Hooft--Feynman
gauge, which contribute to the order
${\cal O}({m_t}^2\log {\Lambda}^2)$.}
\label{fey}
\end{figure}
\begin{figure}
\caption{Two--dimensional projection in the plane of
$\kln$ and $\krn$, for $m_t=150$\,GeV, $m_H=100$\,GeV.}
\label{pr1}
\end{figure}
\begin{figure}
\caption{Two--dimensional projection in the plane of $\kln$ and $\klc$, for
$m_t=150$\,GeV, $m_H=100$\,GeV.}\label{pr2}
\end{figure}
\begin{figure}
\caption{
Two--dimensional projection in the plane of $\krn$ and $\klc$, for
$m_t=150$\,GeV, $m_H=100$\,GeV.}
\label{pr3}
\end{figure}
\begin{figure}
\caption{ The allowed region of
$\kappa_L$ and $\kappa_R$, for
$m_t=150$\,GeV, $m_H=100$\,GeV. (Note that $\kappa_L =\kln=2\klc$
and $\kappa_R =\krc$.)}
\label{f5}
\end{figure}
\begin{figure}
\caption{ The allowed region
 of $\kappa_L$ and $\kappa_R$, for
$m_t=175$\,GeV, $m_H=100$\,GeV. (Note that $\kappa_L =\kln=2\klc$
and $\kappa_R =\krc$.)}
\label{f6}
\end{figure}
\begin{figure}
\caption{ The allowed range of  ($\kappa_L - 2\kappa_R$) as a
function of the mass of the top quark. (Note that $\kappa_L =\kln=2\klc$
and $\kappa_R =\krc$.)}
\label{f7}
\end{figure}
\begin{figure}
\caption{ The allowed range of the coupling $\klc=\kln/2=\kappa_{L}/2$
as a function of the top quark mass.}
\label{f8}
\end{figure}
\begin{figure}
\caption{ The allowed region of  $\kln$
and  $\krn$, for models without a light Higgs boson, and
$m_t=175$\,GeV. }
\label{f9}
\end{figure}
\begin{figure}
\caption{ A comparison between our model and the model in Ref.~[7].
The allowed regions in both models are shown on the plane of
$\kln$ and $\krn$, for $m_{t}=150$\,GeV.}
\label{f10}
\end{figure}
\begin{figure}
\caption{
The allowed
$|\krc|$ and $\klc$ are bounded within
the two dashed (solid) lines
for a 20\% (50\%)  error on the measurement
of the single--top production rate, for a 175 GeV top quark.}\label{f11}
\end{figure}
\begin{table}
\caption{The confined range of the couplings, $\kappa_L$ and $\kappa_R$
for various top masses.}
\begin{tabular}{ccc}
$m_t$ (GeV ) &     $\kappa_L$     &    $\kappa_{R}$  \\ \tableline
150          &   $-0.10$ ----- 0.50   &    $-0.15$ ----- 0.25  \\
175          &   $-0.05$ ----- 0.40   &    $-0.10$ ----- 0.20    \\
200          &   $ 0.0$  ----- 0.35   &    $-0.05$ ----- 0.15    \\
300          &   $ 0.10$ ----- 0.25   &      0.00  ----- 0.10
\end{tabular}
\end{table}


\newpage
\begin{references}
\bibitem{mele}The LEP Collaborations
ALEPH, DELPHI, L3, OPAL and The LEP Electroweak
Working Group, CERN/PPE/93--157 (1993); \\
W. Hollik, {\it talk at the $XVI$ Int. Symp. on Lepton--Photon Interactions,
Cornell University, Ithaca, N.Y., Aug 10-15 (1993)}; \\
M. Swartz, {\it talk at the $XVI$ Int. Symp. on Lepton--Photon Interactions,
Cornell University, Ithaca, N.Y., Aug 10-15 (1993)}; \\
Barbara Mele, {\it Invited talk at the $XIV$ Encontro
Nacional de Fisica de Campos e Particulas, Caxambu, Brazil, 29 Sept-3 Oct,
1993}.
\bibitem{alta}G. Altarelli, CERN--TH--6867/93.
\bibitem{pc}For a discussion see, R.D. Peccei,
{\it Lectures given at the 1993 Scottish
Summer School, St. Andrews, Scotland, August 1993, and at the 1993
Escuela Latino Americano de Fisica, Mar del Plata, Argentina, Jul 1993}
\bibitem{kane}G.L. Kane, UM--TH--91--32, Dec 4,1991, {\it Invited
Lectures at the Workshop on High Energy Phenomenology, Mexico City, July
1--10, 1991}.
\bibitem{D0}S. Abachi {\it et al.}, Phys. Rev. Lett. {\bf 72,} 2138 (1994).
\bibitem{CDF}F. Abe {\it et al.} The CDF Collaboration,
Fermilab--PUB--94--097--E, Apr 1994.
\bibitem{pczh}R.D. Peccei, S. Peris and X. Zhang, Nucl. Phys. {\bf B349,}
305 (1990).
\bibitem{sekh}R.S. Chivukula, E. Gates, E.H. Simmons and J. Terning,
Phys. Lett. {\bf B311,} 157--162 (1993);\\
R.S. Chivukula, E.H. Simmons and J. Terning, BUHEP--94--8, Apr 1994.
\bibitem{lopez}Jorge L. Lopez, D.V. Nanopoulos, Gye T. Park, Xu Wang
and A. Zichichi, CERN--TH--7139/94.
\bibitem{feld}Genevi\'eve Belanger and Gordon Feldman, JHU--HET 8406, Aug
 1984;\\
Gautam Bhattacharyya, CERN--TH--7196/94, Mar 1994.
\bibitem{pesk}Michael E. peskin and Tatsu Takeuchi, Phys. Rev. Lett.
{\bf 65,} 964 (1990);\\
D.C. Kennedy and P. Langacker, Phys. Rev. Lett. {\bf 65,} 2967 (1990);\\
B. Holdem, Phys. Lett. {\bf B259,} 329 (1991);\\
A. Ali and G. Degrassi, DESY 91--035 (1991).
\bibitem{bar}R. Barbieri, IFUP--TH 23/93, July 1993, in {\it Lectures
given at the Symposium on Particle Physics at the Fermi scale, Beijing, May
27 --June 4, 1993}.
\bibitem{bar1}Guido Altarelli, Riccardo Barbieri and Francesco
Caravaglios, Nucl. Phys. {\bf B405,} 3 (1993).
\bibitem{bar2}G. Altarelli and R. Barbieri, Phys. Lett. {\bf B253,}
161 (1990); \\
G. Altarelli, R. Barbieri and S. Jadach, Nucl. Phys. {\bf B369,} 3 (1992).
\bibitem{geor}Howard Georgi, Nucl. Phys. {\bf B361,} 339 (1991).
\bibitem{gold}M. Golden and L. Randall, Nucl. Phys. {\bf B362,} 3 (1991);\\
R.D. Peccei and S. Peris, Phys. Rev. {\bf D44,} 809 (1991); \\
A. Dobado {\it et al.}, Phys. Lett. {\bf B255,} 405 (1991); \\
M. Dugan and L. Randall, Phys. Lett. {\bf B264,} 154 (1991).
\bibitem{buch}W. Buchm\"{u}ller and D. Wyler, Nucl. Phys.
{\bf B268,} 621 (1986).
\bibitem{wwww}For a review, see C.--P. Yuan, {\it published in Perspectives
on Higgs Physics, edited by G. Kane, World Scientific}, 1992, pp 415--428.
\bibitem{tana}Y. Nambu, {\it in Proc. of the 1988 International
Workshop on New Trends in Strong Coupling gauge Theories, Nagoya, Japan,
eds. by M. Bando, T. Muta and K. Yamawaki
(World Scientific,Singapore, 1989)};\\
W.A. Bardeen, C.T. Hill and M. Lindner, Phys. Rev. {\bf D41,} 1647 (1990);\\
Ralf B\"{o}nisch and Arnd Leike, DESY 93--111, {\it August 1993};\\
Bernd A. Kniehl and Alberto Sirlin, DESY 93--194, NYU--TH--93/12/01,
{\it December 1993}.
\bibitem{wein}S. Weinberg, Physica {\bf 96A,} 327 (1979).
\bibitem{geor2}Howard Georgi, Weak Interactions and Modern Particle
Theory (The Benjamin/Cummings Publishing Company, 1984).
\bibitem{cole}S. Coleman, J. Wess and Bruno Zumino, Phys. Rev.
{\bf D177,} 2239 (1969);\\
C.G. Callan, S. Coleman, J. Wess and Bruno Zumino, Phys. Rev.
{\bf D177,} 2247 (1969).
\bibitem{fer}F. Feruglio, Int.J. Mod. Phys. {\bf A8,} 4937 (1993).
\bibitem{nlc}
See, {\it e.g.},
P. Chen, Phys. Rev. {\bf D46,} (1992) 1186; and the references therein.
\bibitem{pecc}R.D. Peccei and X. Zhang, Nucl. Phys. {\bf B337,} 269 (1990).
\bibitem{hol}B. Holdom and J. Terning, Phys. Lett. {\bf 247B,} 88 (1990).
\bibitem{how1}Howard Georgi, Nucl. Phys. {\bf B363,} 301 (1991).
\bibitem{fer1}Ferruccio Feruglio, Antonio Masiero and Luciano Maiani,
Nucl. Phys {\bf B387,} 523 (1992).
\bibitem{app}Thomas Appelquist and Guo--Hong Wu, Phys. Rev. {\bf D48,}
3241 (1993).
\bibitem{burg}C.P. Burgess and David London, Phys. Rev. {\bf D48,}
4337 (1993).
\bibitem{mart}Martin B. Einhorn, UM--TH--93--12, Apr 1993.
\bibitem{velt}M. Veltman, Nucl. Phys. {\bf B123,} 89 (1977).
\bibitem{long}Anthony C. Longhitano, Phys. Rev. {\bf D22,} 1166 (1980).
\bibitem{thesis}Hwanbae Park, SLAC--Report--435, December 1993.
\bibitem{anlrev}
C.--P. Yuan, {\it et al.},
{\it Report of the subgroup on the Top Quark},
Proceedings of Workshop on Physics at Current Accelerators and Supercolliders,
edited by J. Hewett, A. White and D. Zeppenfeld, 1993, pp 495--505;
and the references therein.
\bibitem{kanetop}
G. Kane, G.A. Ladinsky and C.--P. Yuan, Phys. Rev. {\bf D45} (1992) 124.
\bibitem{carlson}
D.O. Carlson and C.--P. Yuan, in preparation.
\bibitem{book}
Review of Particle Properties, Phys. Rev. {\bf D45}, June 1992, p III.66.
\bibitem{gal}
G.A. Ladinsky and C.--P. Yuan, preprint MSUTH 92/07, to appear in
Phys. Rev. {\bf D} (1994).
\end{references}
\end{document}